\documentclass[reprint,aps,superscriptaddress]{revtex4-1}

\usepackage[T1]{fontenc}
\usepackage{mathptmx}
\usepackage{amsmath}
\usepackage{amssymb}
\usepackage[tight]{units}
\usepackage{color,graphicx}
\usepackage[colorlinks,citecolor=blue,urlcolor=blue,linkcolor=blue]{hyperref}
\DeclareMathOperator\erf{erf}
\newcommand{\sce}{\eta}

\begin{document}
\newcommand{\mpi}{\affiliation{Max-Planck-Institut f\"ur Physik, D-80805 M\"unchen, Germany}}
\newcommand{\coimbra}{\affiliation{Departamento de Fisica, Universidade de Coimbra, P3004 516 Coimbra, Portugal}}
\newcommand{\vienna}{\affiliation{Institut f\"ur Hochenergiephysik der \"Osterreichischen Akademie der Wissenschaften, A-1050 Wien, Austria \\ and Atominstitut, Vienna University of Technology, A-1020 Wien, Austria}}
\newcommand{\tum}{\affiliation{Physik-Department, Technische Universit\"at M\"unchen, D-85747 Garching, Germany}}
\newcommand{\tuebingen}{\affiliation{Eberhard-Karls-Universit\"at T\"ubingen, D-72076 T\"ubingen, Germany}} 
\newcommand{\oxford}{\affiliation{Department of Physics, University of Oxford, Oxford OX1 3RH, United Kingdom}}
\newcommand{\wmi}{\affiliation{Walther-Mei\ss ner-Institut f\"ur Tieftemperaturforschung, D-85748 Garching, Germany}}
\newcommand{\lngs}{\affiliation{INFN, Laboratori Nazionali del Gran Sasso, I-67010 Assergi, Italy}}

\author{G.~Angloher}
  \mpi

\author{A.~Bento}
\coimbra 

\author{C.~Bucci}
\lngs 

\author{L.~Canonica}
\lngs 

\author{A.~Erb}
  \tum
  \wmi

\author{F.~v.~Feilitzsch}
\tum 

\author{N.~Ferreiro~Iachellini}
\mpi

\author{P.~Gorla}
\lngs 

\author{A.~G\"utlein}
\vienna

\author{D.~Hauff}
\mpi 

\author{P.~Huff}
\mpi

\author{J.~Jochum}
\tuebingen 

\author{M.~Kiefer}
\mpi

\author{C.~Kister}
\mpi

\author{H.~Kluck}
\vienna

\author{H.~Kraus}
  \oxford

\author{J.-C.~Lanfranchi}
\tum

\author{J.~Loebell}
\tuebingen

\author{A.~M\"unster}
\tum

\author{F.~Petricca}
\mpi 

\author{W.~Potzel}
\tum 

\author{F.~Pr\"obst}
\email{proebst@mpp.mpg.de}
  \mpi

\author{F.~Reindl}
\email{florian.reindl@mpp.mpg.de}
  \mpi

\author{S.~Roth}
\tum 

\author{K.~Rottler}
\tuebingen 

\author{C.~Sailer}
\tuebingen 

\author{K.~Sch\"affner}
\lngs 

\author{J.~Schieck}
\vienna 

\author{J.~Schmaler}
\mpi

\author{S.~Scholl}
\tuebingen

\author{S.~Sch\"onert}
\tum 

\author{W.~Seidel}
\mpi 

\author{M.~v.~Sivers}
\tum 

\author{L.~Stodolsky}
\mpi 

\author{C.~Strandhagen}
  \email{strandhagen@pit.physik.uni-tuebingen.de}
  \tuebingen

\author{R.~Strauss}
\mpi 

\author{A.~Tanzke}
\mpi 

\author{M.~Uffinger}
\tuebingen 

\author{A.~Ulrich}
\tum 

\author{I.~Usherov}
\tuebingen 

\author{S.~Wawoczny}
\tum 

\author{M.~Willers}
\tum 

\author{M.~W\"ustrich}
\mpi 

\author{A.~Z\"oller}
  \tum
%
%
%\institute
%{Max-Planck-Institut f\"ur Physik, D-80805 M\"unchen, Germany \label{addrMPI} \and
%Departamento de Fisica, Universidade de Coimbra, P3004 516 Coimbra, Portugal \label{addrCoimbra} \and
%INFN, Laboratori Nazionali del Gran Sasso, I-67010 Assergi, Italy \label{addrLNGS} \and
%Physik-Department, Technische Universit\"at M\"unchen, D-85747 Garching, Germany \label{addrTUM} \and
%Walther-Mei\ss ner-Institut f\"ur Tieftemperaturforschung, D-85748 Garching, Germany \label{addrWMI} \and 
%Institut f\"ur Hochenergiephysik der \"Osterreichischen Akademie der Wissenschaften, A-1050 Wien, Austria\\
%and Atominstitut, Vienna University of Technology, A-1020 Wien, Austria \label{addrVienna} \and
%Eberhard-Karls-Universit\"at T\"ubingen, D-72076 T\"ubingen, Germany \label{addrTUE} \and
%Department of Physics, University of Oxford, Oxford OX1 3RH, United Kingdom \label{addrOxford}
%}
%
%\thankstext{e1}{corresponding author: proebst@mpp.mpg.de}
%\thankstext{e2}{corresponding author: freindl@mpp.mpg.de}
%\thankstext{e3}{corresponding author: strandhagen@pit.physik.uni-tuebingen.de}

\title{Results on low mass WIMPs using an upgraded CRESST-II detector}

\begin{abstract}
The CRESST-II cryogenic dark matter search aims for the detection of WIMPs via elastic scattering off nuclei in CaWO$_4$ crystals. We present results from a low-threshold analysis of a single upgraded detector module. This module efficiently vetoes low energy backgrounds induced by $\alpha$-decays on inner surfaces of the detector. With an exposure of \unit[29.35]{kg live days} collected in 2013 we set a limit on spin-independent WIMP-nucleon scattering which probes a new region of parameter space for WIMP masses below \unit[3]{GeV/c$^2$}, previously not covered in direct detection searches. A possible excess over background discussed for the previous CRESST-II phase 1 (from 2009 to 2011) is not confirmed. 
\end{abstract}

\maketitle
%Compiled on \today\ at \currenttime
\section{Introduction}
CRESST-II is a cryogenic dark matter search experiment located at the Laboratori Nazionali del Gran Sasso in Italy. Scintillating CaWO$_4$ crystals are used as a multi-element target for the direct search for WIMPs (Weakly Interacting Massive Particles) via elastic scattering off nuclei. Inside a reflective and scintillating housing, each CaWO$_4$ crystal is paired with a light detector measuring the scintillation light. Crystal and light detector are operated as two independent cryogenic calorimeters, each equipped with a superconducting tungsten transition edge sensor (TES) read out by a SQUID, and a heater for controlling the operating temperature and injecting heater pulses. The signal from the TES on the CaWO$_4$ crystal (phonon channel) provides a precise measurement of the energy deposited in the crystal, while the light signal, measured with a TES on the light absorber (light channel), is used for event-type discrimination. The sought-for nuclear recoils are distinguished from the dominant radioactive $e^-/\gamma$-background by their much smaller light signal (quenching). The amount of scintillation light also depends on the mass of the recoiling nucleus. Thus, measuring the scintillation light helps to disentangle recoils off the three different target nuclei (O, Ca and W).

Several experiments, DAMA \cite{Bernabei2013}, CoGeNT \cite{Aalseth2013}, CRESST-II \cite{Angloher2012} and CDMS II Si \cite{Agnese2013} reported observation of an excess of events at low energies above their background estimates. This could be attributed to scattering of light WIMPs with a mass in the 6 to 30 GeV/c$^{2}$ range, while other experiments like XENON100 \cite{Aprile2012}, LUX \cite{Akerib2014} and SuperCDMS \cite{Agnese2014} exclude this region of parameter space.  The main challenge of detecting WIMPs with such low masses is to measure the small recoil energies, of at most a few keV, and still achieve sufficient background discrimination.

To clarify the nature of the signal excess reported in \cite{Angloher2012}, the CRESST collaboration startet CRESST-II phase 2 with upgraded detectors in July 2013. Data from August 2013 to the beginning of January 2014 were used to study the performance of new types of detector modules. Non-blinded \unit[29.35]{kg live days} of data of a single detector module from this first period will be used to derive the low-mass WIMP limit presented in this letter.

\section{Set-up and detector modules}
A detailed description of the CRESST-II set-up, data acquisition (DAQ), readout,  and the procedures for controlling detector stability, as well as reconstructing the deposited energy from the measured pulses can be found in earlier publications~\cite{Angloher2009_run30,Angloher2005_cresstIIproof}. 

Four of the 18 CaWO$_4$ crystals operated in the present phase 2 were grown in a recently established facility within the
CRESST collaboration (TU Munich)~\cite{Erb2013}. Due to improved selection of raw materials and the control of all production steps, these crystals show a factor of 2 to 10 lower $e^-/\gamma$-background in the energy region of interest, as compared to previously available commercial crystals. Also, the level of $\alpha$-contaminations is reduced from \unit[$\sim$15-35]{mBq/kg} for typical commercial crystals to \unit[$\sim$1-3]{mBq/kg} for the ones grown at TU Munich~\cite{muenster2014}. 

The most difficult background in the previous phase were  $^{206}$Pb recoils from $\alpha$-decays of $^{210}$Po on the metal holding clamps of the crystal. These clamps provided the only non-scintillation surface inside the detector housing. Thus, an $\alpha$ absorbed in the clamp could stay undetected. In this letter, we use data from a single detector module, of a new design with fully scintillating inner housing. Instead of metal clamps, CaWO$_4$ sticks fed through holes in the scintillating housing hold the crystal \cite{Strauss2014_StickDesign}. In this design the $\alpha$ from such a decay will always produce additional scintillation light allowing a fully efficient discrimination of such events. A new block-shaped CaWO$_4$ crystal, grown at TU Munich, with a mass of \unit[249]{g} is used. The hardware trigger threshold is set at the energy of \unit[0.6]{keV}.

%The most difficult background in the previous run were $^{206}$Pb recoils from $\alpha$-decays of $^{210}$Po on the holding clamps of the crystal. These clamps provided the only non-scintillating surface inside the detector housing. In this letter, we use data from a single detector module, of a new fully scintillating design. Using CaWO$_4$ sticks, instead of metal holding clamps, this design provides a fully efficient active discrimination of the $^{206}$Pb recoil background \cite{StickDesign}. The hardware trigger threshold is set at the energy of \unit[0.6]{keV}. A new block-shaped CaWO$_4$ crystal, grown at TU Munich, with a mass of \unit[249]{g} is used.
The high temperatures in vacuum needed for the deposition of high-quality tungsten films lead to an oxygen deficit in CaWO$_4$. Such a deficit causes a reduced light output and thus a direct evaporation of the TES on the target crystal should be avoided. Therefore, the tungsten TES is deposited on a separate small CaWO$_4$ carrier which is then glued with epoxy resin onto the large CaWO$_4$ target crystal \cite{Kiefer2009}.

\section{Data set and analysis} \label{sec:Analysis}

\subsection{Energy scale and resolution}\label{sec:EnergyScale}

We calibrate the pulse height response of the two detector channels, phonon and light, to \unit[122]{keV} with \unit[122]{keV} $\gamma$'s from a $^{57}$Co calibration source.
The response of both channels to lower deposited energies is linearized with pulses injected to the heater with a constant rate throughout the run. Combining the information from source and heater pulses then yields the phonon ($E_p$) and light ($E_l$) energy for each event. 
This calibration of the phonon channel implicitly compensates for the fraction of the deposited energy leaving the crystal for \unit[122]{keV} $\gamma$'s (called $\sce$ in the following). 

\begin{figure}[tbp]
  \includegraphics[width=\linewidth]{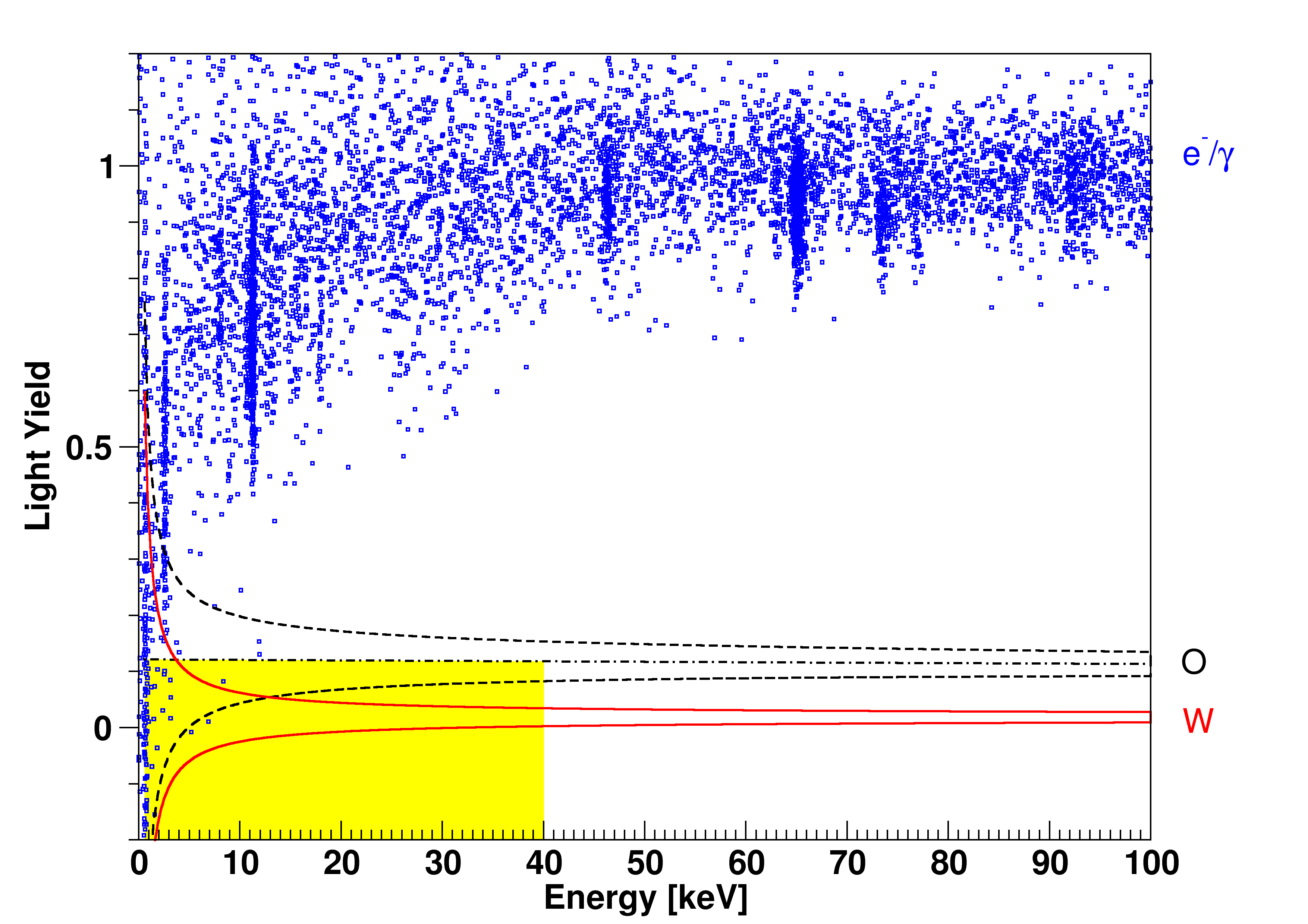}
  \caption{Light yield versus energy of events passing all selection criteria (see Section \ref{sec:TCEff}). The tungsten and oxygen nuclear recoil bands in which we expect the central 80\% of the respective recoils are shown as solid (red) and dashed (black) line. The dash-dotted line marks the center of the oxygen band. Events with energies from \unit[0.6]{keV} to \unit[40]{keV} and light yields below the center of the oxygen band are accepted as WIMP recoil candidates.}
  \label{fig:acceptance_region}
\end{figure}

We define the light yield (LY) as the ratio of both energies ($LY=E_l$/$E_p$). It serves to discriminate different types of interactions. This definition implies a mean light yield of 1 for $\gamma$-events at \unit[122]{keV}. A light yield  $<1$ means that less energy escapes the crystal as scintillation light and more energy remains in the crystal. Thus, the calibration of the phonon channel assigns an energy slightly above the nominal value for such an event.  
The event-type-independent total deposited energy E -- used throughout this letter -- is given by the following relation:
\begin{equation}
								E=\sce E_l + (1-\sce) E_p = [1-\sce(1-LY)]E_p,
                \label{eq:energyCorrection}
\end{equation}
where $\sce$ is the fraction of the deposited energy escaping the crystal as scintillation light for an event with light yield one.

This correction affects events with a light yield <1. These are nuclear recoils and alpha events, but also low energy  $e^-/\gamma$-events, because of the decrease of the light yield of the  $e^-/\gamma$-band towards lower energies (see Figure \ref{fig:acceptance_region}). This decrease can be attributed to a non-proportionality of the light yield, as observed in most inorganic scintillators at low energies \cite{Moses_2008,Lang2009}.

Statistical fluctuations in the amount of scintillation light produced for mono-energetic $\gamma$-events make this correlation visible as a small tilt of the corresponding $\gamma$-lines in the uncorrected energy/light yield-plane. Using this tilt the value of $\sce=0.066\pm0.004$ (stat.) is determined (similar to \cite{Arnaboldi2010}). This correction makes the energy measured for $\alpha$-decays inside the crystal, e.g. those of natural $^{180}$W \cite{Cozzini2004}, consistent with their nominal Q-value. Furthermore, the value determined for $\sce$ is in agreement with dedicated studies on the scintillation efficiency \cite{PhDMichael}.

\begin{figure}[tbp]
\includegraphics[width=\linewidth]{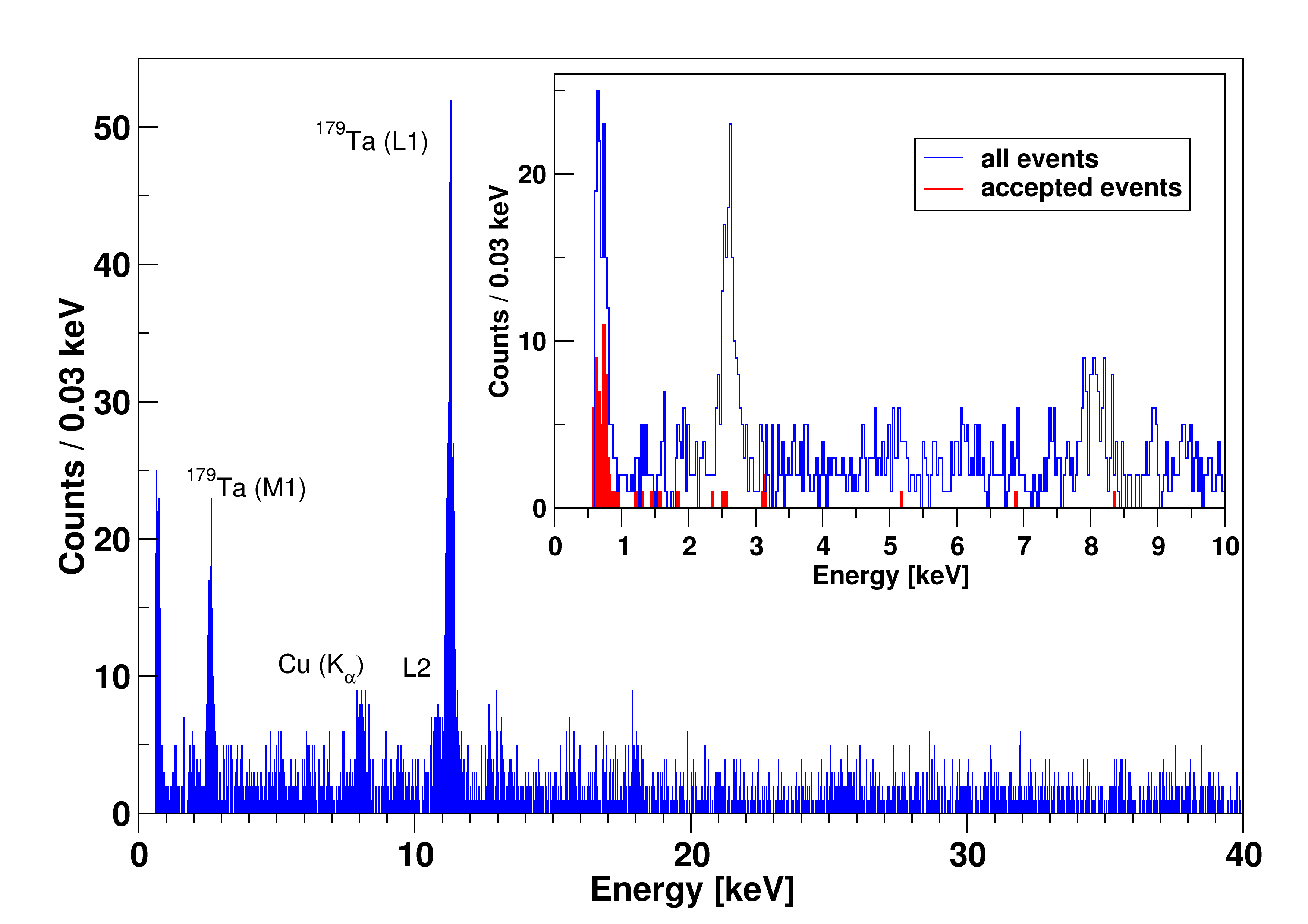}
\caption{
  Low-energy spectrum of all events recorded with a single module and an exposure of \unit[29.35]{kg live days}. The visible lines mainly originate from cosmogenic activation (see text). The insert shows a zoom into the energy spectrum of all events (blue). Shown as filled red histogram are the events in the acceptance region (shaded yellow area in Figure \ref{fig:acceptance_region}).
}
\label{fig:energy_spectrum}
\end{figure}

The resulting energy spectrum of the events in Figure~\ref{fig:acceptance_region} is shown in Figure~\ref{fig:energy_spectrum}. 
The prominent peaks with fitted peak positions of \unit[(2.6014 $\pm$ 0.0108)]{keV} and \unit[(11.273 $\pm$ 0.007)]{keV} can be attributed to M1 and L1 electron capture decays of cosmogenically produced $^{179}$Ta. 
The fitted peak positions agree with tabulated values of \unit[2.6009]{keV} (the binding energy of the Hf M1 shell) and \unit[11.271]{keV} (Hf L1 shell) \cite{firestone} within deviations of \unit[0.5]{eV} and \unit[2]{eV}, respectively. With rather low  statistics an L2 peak is also visible. Its fitted peak position of \unit[(10.77 $\pm$ 0.03)]{keV} also agrees within errors with the tabulated value of \unit[10.74]{keV}. The peak at \unit[(8.048 $\pm$ 0.029)]{keV} is attributed to the copper K$_\alpha$ escape lines. An excellent agreement can also be found at higher energies for the \unit[46.54]{keV} peak of external $^{210}$Pb decays and the \unit[65.35]{keV} peak from K-shell capture decays of $^{179}$Ta. The energy resolution of the peak at \unit[2.601]{keV} is $\Delta$E$_{1\sigma}$=\unit[(0.090 $\pm$ 0.010)]{keV}. With the present trigger setting it could not be clarified, whether the rise towards the threshold energy of \unit[0.6]{keV} is particle-induced, or noise triggers, or both.  All errors quoted are statistical \unit[1]{$\sigma$} errors.

\subsection{Trigger and cut efficiencies}\label{sec:TCEff}
\begin{figure}[tbp]
  \includegraphics[width=\linewidth]{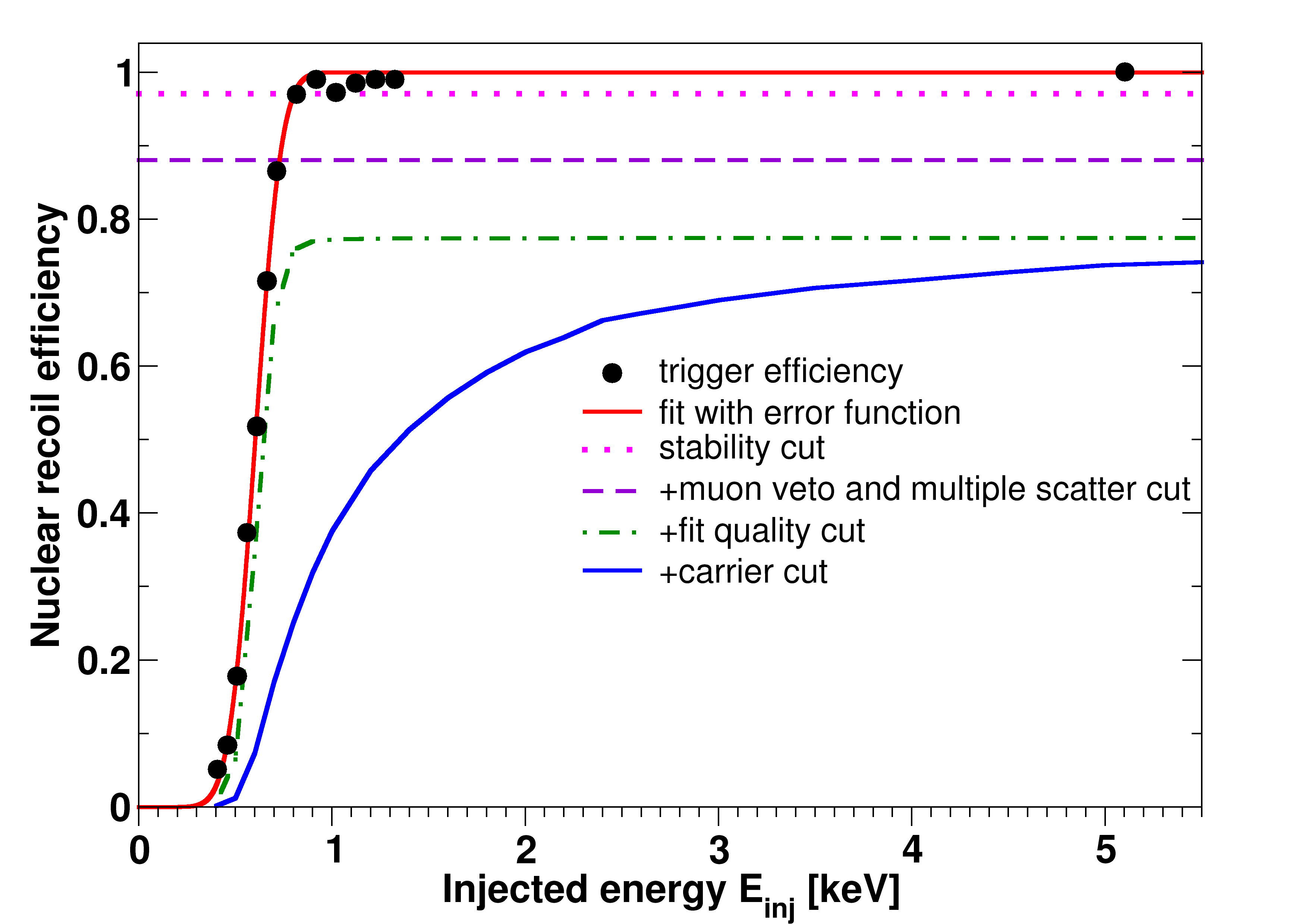}
	\caption
  {
    The filled circles (black) are trigger efficiencies measured by injecting heater pulses with closely spaced discrete energies. The full (red) curve is a fit with an error function which yields an energy resolution (\unit[1]{$\sigma$}) of \unit[(107 $\pm$ 3)]{eV} and an energy threshold of \unit[(603 $\pm$ 2)]{eV}. Also shown in this plot is the nuclear recoil efficiency after  cumulative application of each signal selection criterion as described in the text. The energy $E_{\text{inj}}$ corresponds to an $e^-/\gamma$-event without applying the small correction of equation \ref{eq:energyCorrection}.
}
	\label{fig:efficiencies}
\end{figure}

The trigger efficiency is determined by injecting low energy pulses with the heater. The fractions of heater pulses causing a trigger for each injected energy $E_{\text{inj}}$ are shown as solid circles (black) in Figure~\ref{fig:efficiencies}. Errors are smaller than the symbol size. The energy $E_{\text{inj}}$ is calibrated with \unit[122]{keV} $\gamma$'s (see section \ref{sec:EnergyScale}). The solid curve (red) is a fit with the function  $f(E_{\text{inj}})=1/2\cdot\{1+\erf[(E_{\text{inj}}-E_{\text{th}})/(\sigma \sqrt{2})]\}$, where erf is the Gaussian error function.  $f(E_{\text{inj}})$ describes the probability that an injected energy $E_{\text{inj}}$ is detected as an energy larger than the threshold energy $E_{\text{th}}$. The fit returns $E_{th}$=\unit[(603 $\pm$ 2(stat.))]{eV} and an energy resolution of $\sigma$=\unit[(107 $\pm$ 3(stat.))]{eV}. This resolution agrees with the energy resolution determined for low-energy $\gamma$-peaks, confirming that the resolution of the phonon channel at low energies is entirely determined by the baseline noise.  

We apply a few quality cuts, as discussed below, on the raw data to remove events where a correct reconstruction of the deposited energy cannot be guaranteed. For all cuts energy dependent efficiencies are measured by applying the cuts on a set of artificial nuclear recoil events closely spaced in energy. These artificial pulses are created by superimposing signal templates, obtained by averaging a large number of pulses from the \unit[122]{keV} $^{57}$Co calibration peak, on empty baselines periodically sampled throughout the run. The templates of phonon and light detector are scaled to correspond to a nuclear recoil event of fixed injected energy. Possible pulse shape differences between electron and nuclear recoils are negligible, also confirmed by a neutron calibration. The efficiencies for a certain injected energy $E_{\text{inj}}$ are then given by the fraction of signals passing each cut. Figure~\ref{fig:efficiencies} shows the cumulative energy dependent nuclear recoil efficiency after each selection criterion.

The first cut is the so-called stability cut, only accepting pulses between two stable heater pulses (sent every six seconds) in order to ensure that both channels of a module were fully operational and running stably at their respective operating points at the time of an event.

Events coincident with a signal in the muon veto and/or with signals in any other detector module (dashed purple line) are also rejected, since multiple scatterings are not expected for WIMPs in view of their rare interactions.

Other invalid pulses (e.g. pile-up events and SQUID resets) are rejected mostly by a cut on the RMS deviation (Root  Mean Square) of a fit of the signal template to the measured pulse of the corresponding detector (dash-dotted green line). 

Events in the TES-carrier exhibit a reduced light output compared to events occurring in the main crystal, possibly mimicking nuclear recoil events. They are efficiently discriminated by a cut using the much shorter rise and decay times of the signal in the phonon channel. We optimize the cut to remove the carrier events as efficient as possible. For low energies (<\unit[5]{keV}) the decay and rise time distributions of events in the carrier and in the main crystal overlap, resulting in a decreasing cut efficiency depicted in solid blue. Since this cut is the last one applied, the solid blue line also marks the final nuclear recoil efficiency.  

For the trigger and cut efficiency the small correction ($\sce$=\unit[6.6]{\%}) given by equation \ref{eq:energyCorrection} is not applied, leading to a slight underestimation of the efficiencies and, therefore, to a conservative WIMP limit.

Baseline noise, trigger and cut efficiency are constant in time throughout the
run. The exposure before cuts, with DAQ dead time accounted for, is \unit[29.35]{kg live days} for the module under consideration.

\subsection{Acceptance region}
  
The region in the energy/light yield-plane where one expects a given nuclear recoil is determined by the resolutions of the light and phonon channel and the quenching factor for the given nucleus. This quenching factor describes the light yield reduction compared to an electron of the same energy. Measured values of quenching factors from \cite{Strauss2014_quenching} have been used in this work. In the energy region of interest the energy resolution of the phonon channel is typically much better than that of the light channel. We extract the resolution of the light channel as a function of detected light energy by fitting the $e^-/\gamma$-band in the energy/light yield-plane with a Gaussian of energy dependent center and width. We note that, although the production of scintillation light is governed by Poisson statistics, the Gaussian model assumption is a very good approximation in our region of interest. This is because the $e^-/\gamma$-events produce a sufficiently large number of photons for the Poisson distribution to be well approximated by a Gaussian distribution. For the quenched bands, on the other hand, the resolution is dominated by the Gaussian baseline noise and Poissonian photon statistics plays a minor role.
 
The lower limit of accepted energies is set at \unit[0.6]{keV}, where the trigger efficiency is \unit[50]{\%}. Since no significant WIMP signal is expected for CaWO$_4$ above \unit[40]{keV} (see e.g. \cite{Angloher2012}), we choose this energy as the upper acceptance boundary. Towards low energies the expected number of WIMP-induced events increases exponentially, while on the other hand the finite energy resolution of the light channel leads to an increased leakage of $e^-/\gamma$-events into the nuclear recoil bands. This leakage occurs first for the oxygen band, with the highest light yield of the three nuclear recoil bands. To limit this leakage at very low recoil energies, we choose the center of the oxygen band as the upper light yield bound of the acceptance region. The resulting acceptance region is shown as the yellow-shaded region in Figure~\ref{fig:acceptance_region}. It includes all three kinds of nuclear recoils, \unit[50]{\%} of all O recoils and, depending on energy, a much larger fraction of all Ca and W recoils. 

\begin{figure}[tbp]
  \includegraphics[width=\linewidth]{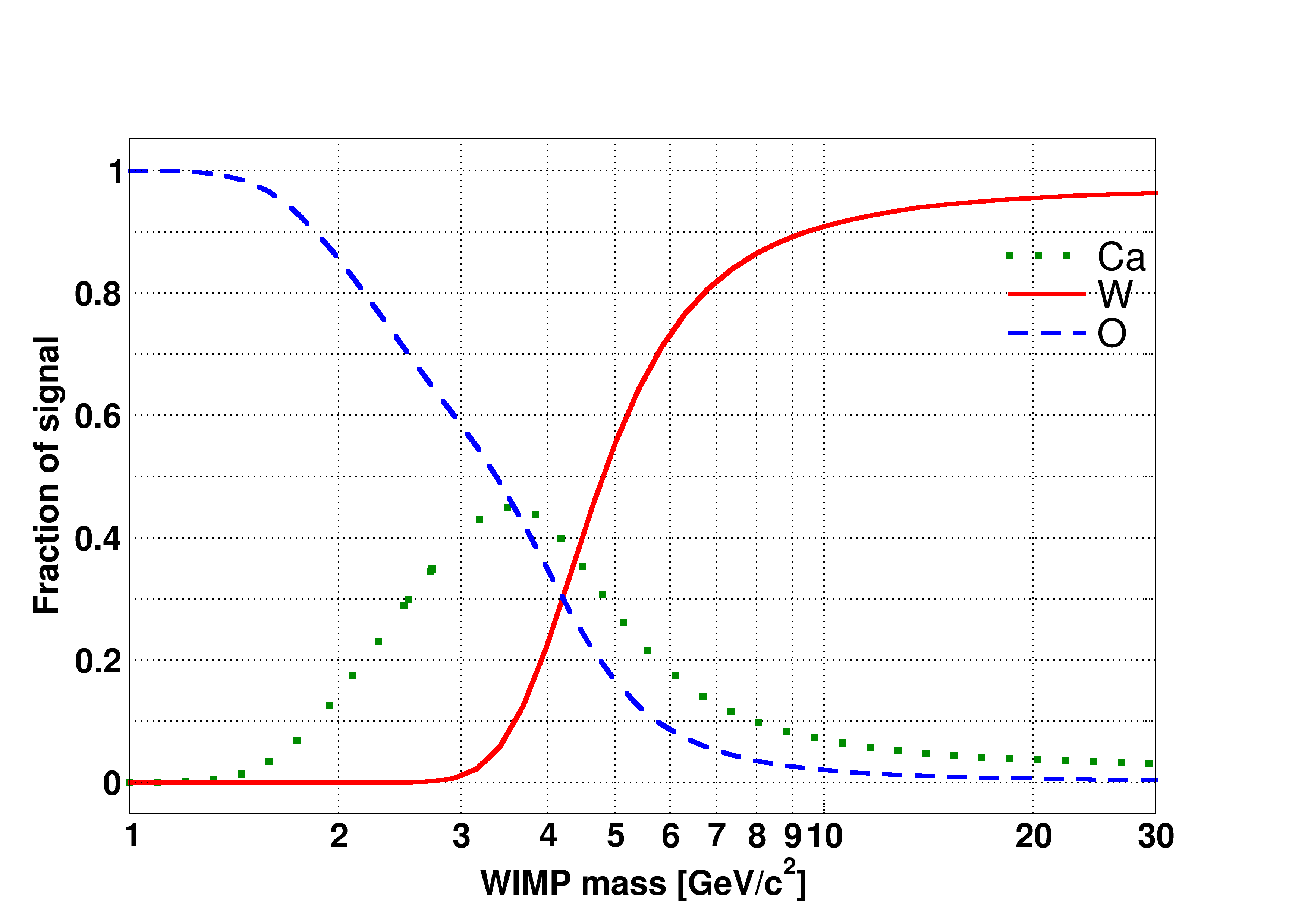}
\caption{Dependence of the fraction of recoils expected for the three different nuclei on the WIMP mass, taking into account the acceptance region shown in Figure~\ref{fig:acceptance_region} and the trigger and cut efficiencies from Figure~\ref{fig:efficiencies}.  }
  \label{fig:composition}
\end{figure}

Depending on the mass of a possible WIMP and the threshold of the detector, any of the nuclei in CaWO$_4$ can be a relevant target for WIMP scattering.  For most WIMP masses, however, the rate of heavy tungsten recoils dominates due to the large coherence factor ($\sim$A$^2$) assumed in the WIMP-nucleon cross section for spin-independent interactions. Only for low WIMP masses, where tungsten recoils are below the energy threshold, the lighter targets calcium and oxygen are important (see Figure~\ref{fig:composition}). Choosing different upper light yield boundaries for the acceptance region was found to have no significant influence on the result of this analysis. This also applies to variations of quenching factors within uncertainties. 

\section{Results and discussion}
Using the data presented in this letter we derive a limit for the cross section of spin-independent WIMP-nucleon scattering, using Yellin's optimum interval method \cite{Yellin2002_Limit}. All  events in the acceptance region (Figure~\ref{fig:acceptance_region}) and their energy distribution (Figure~\ref{fig:energy_spectrum}) are considered.  The expected WIMP recoil spectrum includes the three different target nuclei, the detector resolution, as well as the trigger and cut efficiency (Figure~\ref{fig:efficiencies}).
The Helm form factor \cite{Lewin1996_reviewmathematics} is used to model effects of the nuclear shape. We assume an isothermal dark matter halo with a galactic escape velocity of \unit[544]{km/s}, an asymptotic velocity of \unit[220]{km/s} and a dark matter density of \unit[0.3]{GeV/cm$^{-3}$}. The annual modulation effect is neglected. 

\begin{figure}[tpb]
  \includegraphics[width=\linewidth]{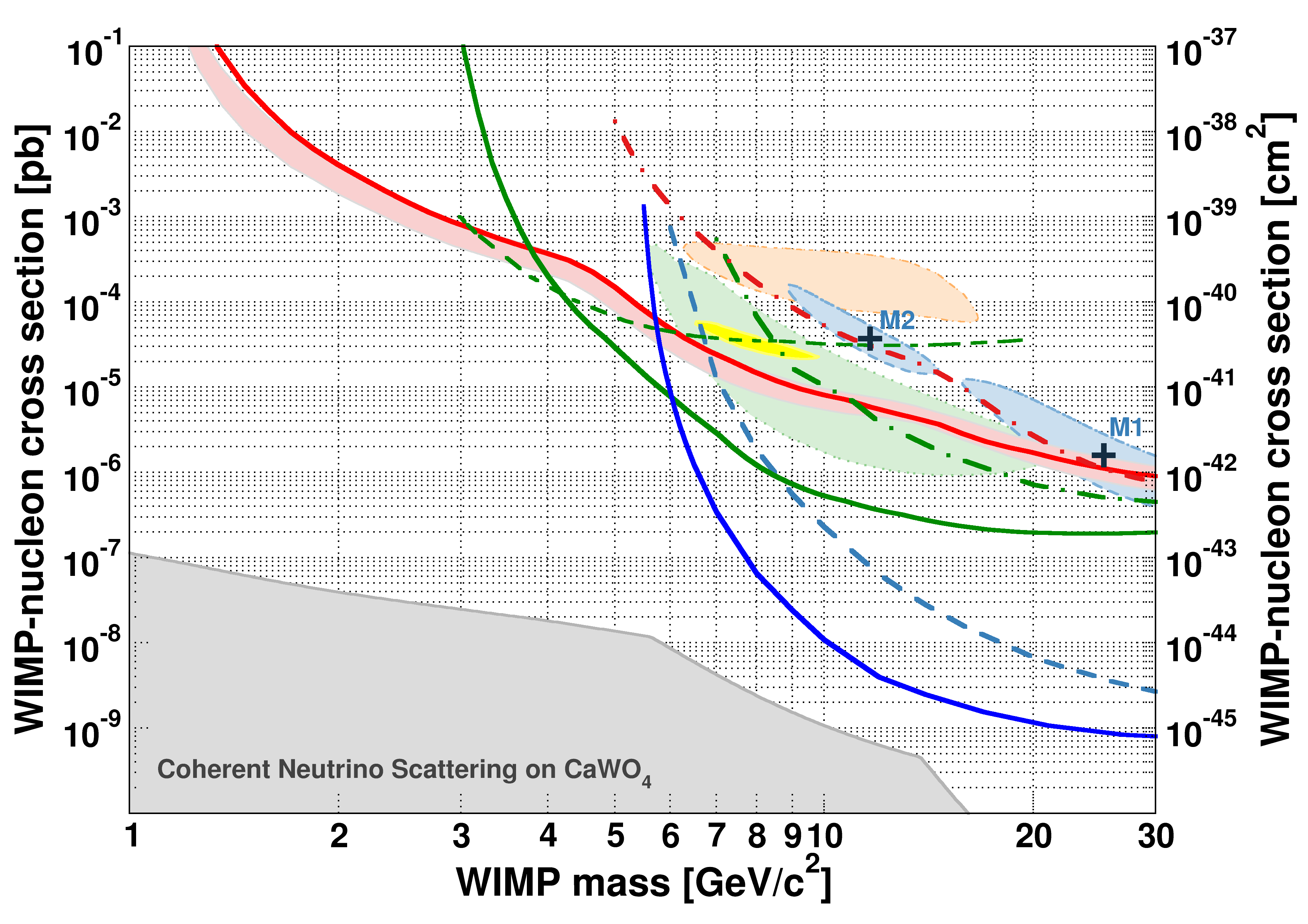}
  \caption
  {
    WIMP parameter space for spin-independent ($\sim$A$^2$) WIMP-nucleon scattering. The \unit[90]{\%} C.L. upper limit (solid red) is depicted together with the expected sensitivity (\unit[1]{$\sigma$} C.L.) from the background-only model (light red band). The CRESST-II \unit[2]{$\sigma$} contour reported for phase 1 in \cite{Angloher2012} is shown in light blue. The dash-dotted red line refers to the reanalyzed data from the CRESST-II commissioning run \cite{Brown2012}. Shown in green are the limits (\unit[90]{\%} C.L.) from Ge-based experiments: SuperCDMS (solid)\cite{Agnese2014}, CDMSlite (dashed) \cite{Agnese_2013SuperCDMS} and EDELWEISS (dash-dotted) \cite{Armengaud2012}. The parameter space favored by CDMS-Si \cite{Agnese2013} is shown in light green (\unit[90]{\%} C.L.), the one favored by CoGeNT (\unit[99]{\%} C.L. \cite{Aalseth2013}) and DAMA/Libra (\unit[3]{$\sigma$} C.L. \cite{Savage2009_DAMAcompatibility}) in yellow and orange. The exclusion curves from liquid xenon experiments (\unit[90]{\%} C.L.) are drawn in blue, solid for LUX \cite{Akerib2014}, dashed for XENON100 \cite{Aprile2012}. 
		Marked in grey is the limit for a background-free CaWO$_4$ experiment arising from coherent neutrino scattering, dominantly from solar neutrinos \cite{Guetlein2014}.
}
\label{fig:msigmaplane}
\end{figure}

The exclusion limit we get is shown as solid red line in Figure~\ref{fig:msigmaplane}. Consistent results were obtained with independent analysis chains (from raw data to final result) based on different software packages.

A Monte Carlo simulation, based on a backround model assuming the presence of e$^-$/$\gamma$-backgrounds only \cite{Strauss2014_Bck}, gives the light-red band (\unit[1]{$\sigma$} C.L.). The limit derived from data and this simulation agree throughout the whole WIMP mass range indicating that the events in the acceptance region may be solely  explained by leakage from the e$^-$/$\gamma$-band. The rise in the energy spectrum below \unit[1]{keV} (see Figure \ref{fig:energy_spectrum}) is not considered in the background model explaining the small difference between simulation and data for WIMP masses smaller than \unit[2]{GeV/c$^2$}. 

The distinctive feature of CRESST-II detectors, to simultaneously probe a potential WIMP signal on light nuclei (O and Ca) in addition to the heavy W nuclei, leads to a more moderate rise of the exclusion limit towards lower WIMP masses when compared with other experiments. The kink at \unit[5]{GeV/c$^2$} marks the transition region from the expected signal rate being dominated by recoils on O and Ca below and on W above this mass (see Figure~\ref{fig:composition}). 

The result presented in this letter clearly excludes the lower mass maximum (M2) of the previous phase \cite{Angloher2012}. More statistics is required to improve our limit at higher WIMP masses and, thus, to clarify the nature of the higher mass maximum (M1). This will be the subject of a blind analysis of additional data collected during the currently ongoing CRESST-II phase 2. 

The improved performance of the upgraded detector manifests itself in a significantly improved sensitivity of CRESST-II for very low WIMP masses. This can be seen by comparing the current limit (solid red line) using the data of a single detector to the one obtained from the reanalyzed commissioning run data (dash-dotted red line) \cite{Brown2012}. For WIMP masses below \unit[3]{GeV/c$^2$} CRESST-II probes new regions of parameter space, previously not covered by other direct dark matter searches.

The sensitivity for light WIMPs can be improved in future runs by further reducing the background level and enhancing the detector performance. Such improvements are realistic and substantial gains in sensitivity for low WIMP masses are possible, even with a moderate target mass.

\section*{Acknowledgements}

This work was supported by funds of the German Federal Ministry of Science and Education (BMBF),
the Munich Cluster of Excellence (Origin and Structure of the
Universe), the Maier-Leibnitz-La\-bo\-ra\-to\-ri\-um (Garching), the Science
and Technology Facilities Council (STFC) UK, as well as the Helmholtz Alliance for Astroparticle Physics. We gratefully acknowledge the work of Michael Stanger from the crystal laboratory of the TU Munich. We are grateful to LNGS for their generous support of CRESST, in particular to Marco Guetti for his constant assistance.

\bibliographystyle{h-physrev}
\bibliography{run33.bib}
%\printbibliography
\end{document}